\documentstyle[epsf,graphics,a4,12pt]{article}
\begin{document}
\newcommand{\beq}{\begin{equation}}
\newcommand{\eeq}{\end{equation}}
\newcommand{\beqn}{\begin{eqnarray}}
\newcommand{\eeqn}{\end{eqnarray}}
\newcommand{\dpf}{\displaystyle\frac}
\newcommand{\no}{\nonumber}
\newcommand{\ep}{\epsilon}

\begin{center}
{\Large Dynamics and holographic discreteness of tachyonic inflation} 
\end{center}
\vspace{1ex}
\centerline{\large 
Bin Wang$^{a,}$\footnote[1]{e-mail:binwang@fudan.ac.cn},
\ Elcio Abdalla $^{b,}$\footnote[2]{e-mail:eabdalla@fma.if.usp.br},
 and Ru-Keng Su $^{c,a}$\footnote[3]{e-mail:rksu@fudan.ac.cn}  
}
\begin{center}
{$^{a}$ Department of Physics, Fudan University, Shanghai 200433,
P. R. China \\
$^{b}$ Instituto de Fisica, Univ. Sao Paulo,
C.P.66.318, CEP
05315-970, Sao Paulo, Brazil \\
$^{c}$ China Center of Advanced Science and Technology (World Lab), Beijing 100080, P. R. China
}
\end{center}

\vspace{6ex}
\begin{abstract}
We consider issues related to tachyonic inflation with inverse-power-law potential. We find the solution of the evolution equations in the slow roll limit in FRW  as well as in the brane cosmology. Using the holographic entropy bound, we estimate the quantum-gravitational discreteness of tachyonic inflation perturbations.
\end{abstract}
\vspace{6ex} \hspace*{0mm} PACS number(s): 98.80.Cq, 98.80.Hw.
\vfill
\newpage
The theory of cosmic inflation offers an attractive resolution of some of the great paradoxes in cosmology and has gained fairly good support from the recent observations of the cosmic microwave background radiation \cite{smoot}. However the accelerated expanding universe cannot be obtained from fundamental theories such as M/String theory. Recently there has been significant progress in bringing about a reconciliation, particularly due to Sen, who has pointed out that rolling tachyons can contribute a mass density to the universe that resembles classical dust \cite{sen}. This has brought a new understanding of the role of the tachyon in string theory. The process of rolling tachyon in the cosmological background has been investigated in [3-11]. It is argued that the tachyonic field might play the dual role of inflation at the early epochs and of some new form of dark matter at late times. Nevertheless some problems with identifying the tachyon as the inflation are discussed in \cite{linde}, where it was indicated that the reheating mechanism is lacking in these models.

Recently an interesting development in cosmology has been introduced, which is called a brane world based on superstring or M theory \cite{dvali}\cite{rs}. In a brane world scenario, our universe is embedded in higher dimensions and standard-model particles are confined to four-dimensional hypersurfaces, while gravity is propagating in higher dimensions. This scenario suggests that the gravitational law could be different from the conventional cosmological model at the early stage of the universe or at high-energy scale. Accounts of reheating in the brane assisted inflation of non-tachyonic field have been discussed in \cite{lidsey}. Effects of tachyon in context of brane world cosmology have been investigated [17-19,8]. It is interesting to note that arguments of possible solution of the  reheating  problem in the tachyonic inflation have been raised in the brane inflationary scenario [17£¬18].

The general functional form of tachyonic potential is not known. However it is expected that the potential $V(\phi)$ has an unstable maximum at $\phi=0$ and a minimum at $\phi=\infty$. Sen \cite{sen} has shown that the choice of an exponential potential for the tachyonic field satisfies these requirements and leads to some interesting properties. Some other potential forms have been constructed leading to the desired cosmological evolution \cite{pa}\cite{semi2}. In this paper we discuss issues of inflation with tachyon rolling on an inverse-power-law potential $V(\phi)=\mu^{\alpha+4}\phi^{-\alpha}$, which satisfies the tachyonic potential requirements. A scalar field with this specific potential form has been shown with interesting behavior in the quintessence scenario \cite{mizu}. We will  find exact solutions of slow roll equations in the usual four-dimensional FRW cosmology as well as on the brane.

The second motivation of the present paper is to study the perturbation and anisotropy of the tachyonic inflation. They are by far the most important aspect of inflation and are believed to be the ancestors of present cosmological structures. Recently there has been an intriguing idea of studying this problem by using the cosmic holographic principle \cite{bin}. It is found that in fundamental theories where the holographic entropy bound arises from discrete fundamental eigenstates, the amount of inflation in the anisotropy is remarkably limited, which can be described with only about $10^5$ bits per sky-harmonic mode implying that the perturbations should be pixilated in some way \cite{hogan}. In this paper, we are going to extend their discussion to the tachyonic inflation.

In the conventional cosmology the Friedmann equation takes the form
\beq   \label{eq(1)}
H^2=\dpf{\rho}{3M^2 _p}
\eeq
We consider that the matter fields consist of tachyonic field $\phi$ as well as radiational fluids, i.e. $\rho=\rho_{\phi}+\rho_r$. According to Sen \cite{sen}, a rolling tachyon can be described by a fluid with energy density and pressure
\beq \label{eq(2)}
\rho_{\phi}=\dpf{V(\phi)}{\sqrt{1-\dot{\phi}^2}}
\eeq
\beq  \label{eq(3)}
P_{\phi}=-V(\phi)\sqrt{1-\dot{\phi}^2}
\eeq
$V(\phi)$ is the tachyon potential. The equation of motion of the tachyonic field minimally coupled to gravity is 
\beq  \label{eq(4)}
\dpf{V\ddot{\phi}}{1-\dot{\phi}^2}+3HV\dot{\phi}+\tilde{V}=0
\eeq
where $\tilde{V}=\dpf{dV}{d\phi}$. For the radiation fluid, we have
\beq \label{eq(5)}
\dot{\rho}_r+4H\rho_r=0.
\eeq

For a Robertson-Walker tachyon matter dominated universe, the condition to have accelerating expansion is 
\beqn \label{eq(6)}
\dpf{\ddot{a}}{a} & = & H^2+\dot{H}=-\dpf{1}{6M_p^2}(\rho_{\phi}+3P_{\phi}) \no \\
                  & = & -\dpf{1}{6M_p^2}\dpf{V}{\sqrt{1-\dot{\phi}^2}}(3\dot{\phi}^2-2)>0
\eeqn
which requires that 
\beq   \label{eq(7)}
\dot{\phi}^2<2/3.
\eeq
This is different from the condition obtained in the inflation caused by a normal scalar field where $\dot{\phi}^2<V(\phi)$.

In order to have a sufficiently long period of inflation, the tachyonic field should start rolling with small initial $\dot{\phi}$. Inserting our potential form into (\ref{eq(4)}) and assuming a slow rolling condition, we find that 
\beq  \label{eq(8)}
\phi=[(2-\alpha/2)\alpha(\mu^{\alpha+4})^{-1/2}M_p(t-t_0)/\sqrt{3}]^{2/(4-\alpha)}
\eeq
where $t_0$ is an integration constant and we consider it as the moment of starting of the inflation. The universe expands as 
\beq \label{eq(9)}
a=a_0\exp[H_0(t-t_0)^{(4-2\alpha)/(4-\alpha)}],
\eeq
where a constant $H_0$ is given by 
\beq \label{eq(10)}
H_0=\dpf{(\mu^{\alpha+4})^{(2+\alpha)/4}}{(\sqrt{3})^{(2-\alpha)/2}M_p^{(2+\alpha)/2}}\alpha^{-\alpha/2}2^{\alpha/2}\dpf{(4-\alpha)^{(2-\alpha)/2}}{4-2\alpha}.
\eeq
To have the expansion of the universe, we need $\alpha<2$ or $\alpha>4$.

Another quantity of interest in inflation is the number of e-folds during the inflation phase,
\beq  \label{eq(11)}
N(\phi)=\ln\dpf{a(t_{end})}{a(t)}=\int^{t_{end}}_t Hdt=-\int^{\phi_{end}}_\phi \dpf{3H^2V}{\tilde{V}}d\phi.
\eeq
Using the inverse-power-law potential form,
\beqn  \label{eq(12)}
N(\phi) & = & \dpf{\mu^{\alpha+4}}{\alpha M_p^2 (2-\alpha)}\phi^{2-\alpha}\vert^{\phi_{end}}_{\phi} \no \\
& = & \dpf{\mu^{\alpha+4}}{\alpha M_p^2(2-\alpha)}[(4-\alpha)\alpha\mu^{-(\alpha+4)/2}M_p/(2\sqrt{3})]^{(4-2\alpha)/(4-\alpha)}(t-t_0)^{(4-2\alpha)/(4-\alpha)}\vert^{t_{end}}_{t_0} \no \\
& = & \dpf{\alpha}{2(2-\alpha)}
\eeqn
where we considered that the inflation stops when $\dot{\phi}^2\geq 2/3$. To overcome the horizon problem and bring us a suitable flat spectrum of perturbations, we require $N\geq 50-60$ e-folds of inflation. Thus $\alpha$ should be in the range $0<\alpha<2$. 

In order to study the dynamics of the tachyonic field, we also need to discuss the radiation dominated era. Employing (\ref{eq(5)}), $\rho_r\sim a^{-4}$. The expansion law of the universe is $a\propto t^{1/2}$. From the equation of motion for the tachyonic field, we find $\phi\propto t$ and $\rho_{\phi}\propto a^{-2\alpha}$. The density parameter of the tachyonic field is 
\beq \label{eq(13)}
\Omega_{\phi}=\dpf{\rho_{\phi}}{\rho_r+\rho_\phi}\sim \dpf{\rho_\phi}{\rho_r}\sim a^{2(2-\alpha)}.
\eeq
In the reasonable range of the parameter $\alpha$, $0<\alpha<2$, we find that the tachyonic field energy decreases slower than the radiation  energy, and eventually the tachyonic field dominates the radiation. This result can also be got from 
\beq \label{eq(14)}
\dot{\Omega}_{\phi}=\dpf{H\rho_r V(4-3\dot{\phi}^2)}{\sqrt{1-\dot{\phi}^2}\rho^2}.
\eeq
Since $\dot{\phi}_{max}=1$, the density parameter of a tachyon field will never decrease in time. Eq. (\ref{eq(14)}) is not dependent on the potential form. This agrees to the study of tachyonic inflation with exponential potential \cite{sami} and provides support to the analysis in \cite{linde}.

We are going to extend our study to the brane world scenario. We will analyze the dynamics of a tachyonic field in the Randall Sundrum¡¯s second brane scenario \cite{rs}. The discussion here may also be valid in other type of brane world models, in which a quadratic term of energy momentum tensor generically appears. The effective Friedmann equation is described as
\beq \label{eq(15)}
H^2+\dpf{k}{a^2}=\dpf{1}{3}\Lambda^{(4)}+\dpf{\rho}{3m_4^2}+\dpf{\rho^2}{36m_5^6}+\dpf{C}{a^4},
\eeq
where $m_4$ and $m_5$ are four and five dimensional Planck mass, respectively, which satisfies $m_4>>m_5$. $k$ is a curvature constant and $C$ is a constant which transmits bulk graviton influence onto the brane. For simplicity we take $\Lambda^{(4)}, k$ equal to zero in the following discussion.

As for matter field on the brane, we consider tachyonic field $\phi$ as well as the conventional radiation fluid, $\rho=\rho_{\phi}+\rho_r$. We consider only the early stage of the universe, at which the quadratic term is dominant and then matter fluid can be ignored. 

In the tachyonic field dominated era, the Friedmann equation becomes
\beq   \label{eq(16)}
H=\dpf{1}{6m_5^3}\rho_{\phi}=\dpf{1}{6m_5^3}\dpf{V(\phi)}{\sqrt{1-\dot{\phi}^2}}
\eeq
The energy and momentum of each field on the brane are conserved. For the tachyonic field we have
\beq  \label{eq(17)}
\dot{\rho_{\phi}}=-3H(\rho_{\phi}+P_{\phi})
\eeq
which is equivalent to the equation of motion described by (4). 

The condition to accommodate the accelerating expansion in this brane world becomes
\beq \label{eq(18)}
\dpf{\ddot{a}}{a}=\dpf{V^2}{6m_5^6}\dpf{1-3\dot{\phi}^2}{6(1-\dot{\phi}^2)}>0
\eeq
which leads to
\beq \label{eq(19)}
\dot{\phi}^2<1/3.
\eeq
This condition is different from that of the tachyonic inflation in the conventional FRW cosmology.

Analogously to (\ref{eq(8)}), substituting our inverse-power-law potential, the tachyonic field is obtained
\beq \label{eq(20)}
\phi=[2\alpha(2-\alpha)m_5^3/\mu^{\alpha+4}]^{1/(2-\alpha)}(t-t_0)^{1/(2-\alpha)}
\eeq
The scale factor is given by 
\beq  \label{eq(21)}
a=a_0\exp[H_0(t-t_0)^{(2-2\alpha)/(2-\alpha)}]
\eeq
where 
\beq   \label{eq(22)}
H_0=\dpf{1}{6m_5^3}\mu^{\alpha+4}[2\alpha(2-\alpha)m_5^3/\mu^{\alpha+4}]^{-\alpha/(2-\alpha)}\dpf{2-\alpha}{2-2\alpha}
\eeq
In order to have the exponentially expanding universe, $\alpha$ has to be in the range $0<\alpha<1$. This requirement is consistent with that of the reasonable number of e-foldings in the inflation similar to that got in (\ref{eq(12)})
\beq   \label{eq(23)}
N=\dpf{\alpha}{2(1-\alpha)}
\eeq
in the brane inflational world. The reasonable range of taking $\alpha$ is tighter than that in the usual FRW cosmology.

In the radiation dominated era, similar to the previous discussion $\rho_r\sim a^{-4}$. Due to the quadratic density term in the Friedmann equation, $a\propto t^{1/4}$ in the brane scenario. The equation of motion for the tachyonic field tells us that $\phi\propto t$, which results in $\rho_{\phi}\propto a^{-4\alpha}$. Thus the density parameter of the tachyonic field $\Omega_{\phi}\sim a^{4(1-\alpha)}$. For the reasonable values of $\alpha$, the tachyonic field energy density far exceeds the density in the radiation. This behavior obtained in the brane cosmology is qualitatively the same as that in the usual FRW cosmology and also supports the argument in \cite{linde}.

Although facing the reheating problem, tachyonic inflation might still be a possible responsible for an earlier stage of inflation followed by another inflationary era driven by other fields \cite{linde}. The existence of the tachyonic inflation stage may be important for the resolution of major cosmological problems such as the homogeneity, flatness and isotropy problems. Thus it is of interest to investigate the perturbation and anisotropy of the tachyonic inflation.

The maximum observable total entropy and the de Sitter expansion rate are related by $S_{max}=\pi M_p^2/H^2$ in holography. According to \cite{hogan}, we assume that this relation arises from discrete elements so that changes in $S$ and $H$ occurs by integer steps. The tachyonic field $\phi$ is also fixed in a finite set of discrete values, since it determines the tachyonic perturbation which in turn fixes $H$ when the perturbation freezes out.

The COBE normalized value for the amplitude of scalar density perturbations is
\beq  \label{eq(24)}
\delta_H^2=\dpf{\pi}{75}P_{\phi}\approx 1.1\times 10^{-10}
\eeq
where
\beq  \label{eq(25)}
P_{\phi}^{1/2}\approx \dpf{H^2}{2\pi\vert\dot{\phi}\vert\sqrt{V}}
\eeq
for the tachyonic field \cite{noh}.

In the conventional FRW cosmology, we find 
\beq  \label{eq(26)}
P_{\phi}^{1/2}=\dpf{3V^2}{2\pi(3M_p^2)^{3/2}\tilde{V}}
\eeq
during the slow roll inflation.

The steady increase in observable entropy is at a rate
\beq  \label{eq(27)}
\dot{S}_{max}=M_p^2H[\dpf{\pi{\tilde{V}}^2}{3H^4V^2}]=H\dpf{1}{4\pi P_{\phi}}
\eeq
Combing (24), in every inflationary e-folding $S_{max}$ increases by an amount of order $10^8$ just due to the classical evolution of the system.

Now consider the effect of a quantum perturbation in the inflation field of amplitude $\delta\phi$, assuming that it behaves like ``frozen¡± in classical spacetime background, the jump in total entropy associated with a horizon-scale perturbation $\delta\phi$ can be expressed in the form
\beq   \label{eq(28)}
\delta S_{tot}=-2\pi M_p^2\dpf{\delta H}{H^3}=\dpf{(3M_p^2)^{3/2}\pi \tilde{V}}{3V^2}V^{1/2}\dpf{\delta\phi}{H}=\dpf{V^{1/2}}{2P_{\phi}^{1/2}}\dpf{\delta\phi}{H}\approx 10^{4}V^{1/2}\dpf{\delta\phi}{H}
\eeq
where the slow roll condition has been adopted. The quantity $\dpf{\delta\phi}{H}$ can be determined statistically in order $1/M_p^2$ \cite{hogan}, where we have introduced the dimension here. The corresponding increment $\delta S_{qft}\approx 10^4 V^{1/2}/M_p^2$ is roughly the jump in the total observable cosmological entropy associated with the creation or destruction of a simple horizon-scale inflaton quantum. Due to the special form of the equation of motion of the tachyonic field, the information in a typical horizon-scale perturbation depends on the value of $V$ here.

The scale of the inflaton perturbations to be observed in the pattern of microwave anisotropy on the sky including direct images created by the Sachs-Wolfe effect is about $10^4$ \cite{hogan}. To have the observable effect, we need $\delta S_{qft}\geq 10^4$ which requires $V^{1/2}>M_p^2$. Obviously this requirement is impossible to be satisfied. The COBE measurement of the CMB anisotropy tells us that when our universe leaves the horizon the potential $V(\phi)$ is far below the Planck scale. To be precise $V^{1/4}$ is no more than a few times $10^{16}Gev$ $(V^{1/4}<M_p)$ and it may be many orders of magnitude smaller \cite{riotto}. This reasonably leads the amount of information $\delta S_{qft}$ attached to the perturbation in the horizon-size inflaton quantum $\delta\phi$ as it freezes into a classical state to be much smaller than the total growth of information of the spacetime during the same time. Therefore the discreteness due to the tachyonic inflation in usual FRW cosmology do not have the observational impact.

Now let us discuss the tachyonic inflation in the brane world. Using the slow roll condition and the Friedmann equations, we have 
\beq  \label{eq(29)}
P_{\phi}^{1/2}=\dpf{3V^{7/2}}{2\pi(6m_5^3)^3\tilde{V}}
\eeq
The maximum observable total entropy is $S_{max}=m_4^2\pi/H^2$ in the brane world. The total growth of information of the spacetime reads
\beqn   \label{eq(30)}
\dot{S}_{max} & = & -\dpf{2m_4^2\pi}{H^3}\dpf{dH}{dt}=H\dpf{\pi m_4^2}{9m_5^3}(6m_5^3)^5\dpf{{\tilde{V}}^2V}{V^7} \no \\
& = & H\dpf{m_4^4V}{24\pi m_5^6P_{\phi}}\approx 10^7\dpf{m_4^2V}{m_5^6}H
\eeqn
The change of the total observable entropy due to $\delta\phi$ has the form
\beqn   \label{eq(31)}
\delta S_{tot} & = & -\dpf{2\pi m_4^2\delta H}{H^3}=\dpf{\pi m_4^2}{3m_5^3}[\dpf{\delta\phi}{H}](6m_5^3)^2\dpf{\tilde{V}}{V^2} \no\\
& = & \dpf{m_4^2}{12m_5^6}[\dpf{\delta\phi}{H}]\dpf{V^{3/2}}{P_{\phi}^{1/2}}\approx \dpf{V^{3/2}}{m_5^6}10^4
\eeqn
where $[\dpf{\delta\phi}{H}]$ has been determined statistically as $1/m_4^2$ \cite{hogan}. The ratio of the amount of information to the total growth of information of the spacetime is $\dpf{V^{1/2}}{m_4^2 10^3}$, which is reasonably much smaller than unit since $V^{1/4}<m_4$. Considering $V^{1/4}=6.2\epsilon^{1/4}\times 10^{16} Gev$ when the observable universe leaves the Hubble parameter, and $m_5\ll m_4=10^{18}Gev$, the corresponding increment $\delta S_{qft} >10^4$ may be satisfied. This may produce a detectable level of discretness.

Suppose that the possible values of the horizon size inflaton perturbation $\delta\phi$ are selected from a discrete set of levels separated by some size $\Delta\phi$. The incremental observable information available by the transition between levels is 
\beq   \label{eq(32)}
\Delta S_{tot}=\dpf{m_4^2\pi}{3m_5^3}[\dpf{\Delta\phi}{H}]\dpf{\tilde{V}}{H^2}
\eeq
We now assert that jumps in $\delta\phi$ occur in steps of at least a certain minimum size $\Delta \phi$, such that $\Delta S_{tot}\geq\ln 2$. This leads to 
\beq   \label{eq(33)}
[\dpf{\Delta\phi}{H}]\geq \dpf{3m_5^3}{m_4^2\pi}\dpf{H^2}{\tilde{V}}\ln 2 >10^{-4}\ln 2
\eeq
where the quantum transition in the maximum observable entropy $\Delta S_{qft}>10^4$ is taken to have observable effect. According to \cite{hogan}, the amplitude of the anisotropy on the sky coming from a discrete distribution with values of the temperature is 
\beq   \label{eq(34)}
\Delta T\geq [\dpf{\Delta\phi}{H}]<Q>\approx 10^{-9}
\eeq
where $<Q>\approx 18\pm 1.6\mu k$ from the COBE data.

In summary, we have studied tachyonic inflation with inverse-power-law potential and found the exact solution of evolution equations in the slow roll limit in the FRW cosmology as well as on the brane. Some differences in the dynamics of tachyonic field have been found between the brane and the conventional cosmological models. We have also shown that the tachyonic matter dominates always in the inflation with the inverse-power-law potential, in agreement with the discussion of exponential potential \cite{sami} and analysis in \cite{linde}. The reheating problem arises both in the usual FRW and brane world. By using the holographic entropy bound, we have estimated the quantum-gravitational discretness of tachyonic inflationary perturbations. In the FRW world, there is no observational effect of the discretness, however in the brane scenario, the discretness may be detectable. The level of discretness in the brane world is calculated. It is worth noting that the discretness estimation here do not depend on the detailed forms of the potential.

\end{document}